\documentstyle[12pt]{article}
\newcommand{\be}{\begin{equation}}
\newcommand{\ee}{\end{equation}}
\newcommand{\ba}{\begin{array}{rcl}}
\newcommand{\ea}{\end{array}}
\begin{document}
\begin{center}
{\Large Correlation between Correlations:\\
Process and Time in Quantum Networks}\\
\medskip
G\"unter~Mahler and Ilki~Kim\\
Institut f\"ur Theoretische Physik~I, Universit\"at Stuttgart\\
Pfaffenwaldring 57, 70550 Stuttgart, Germany\\
{\it email}: mahler@theo.physik.uni-stuttgart.de\\
\end{center}

%%%%%%%%%%%%%%%%%%%%%%%%%%%%%%%%%
\begin{abstract}
 We study a 
 special inhomogeneous quantum network consisting of a ring of $M$ 
 pseudo-spins (here $M = 4$) sequentially coupled to one and the same 
 central spin under the influence of given pulse sequences (quantum 
 gate operations). This architecture could be visualized as a quantum 
 Turing machine with a cyclic ``tape''. Rather than input-output-relations 
 we investigate the resulting process, i.e. the correlation 
 between one- and two-point expectation values 
 (``correlations'') over various time-steps. The resulting 
 spatio-temporal pattern exhibits many non-classical features including 
 Zeno-effects, violation of temporal Bell-inequalities, and quantum 
 parallelism. Due to the strange web of correlations being built-up, 
 specific measurement outcomes for the tape may refer to one or several 
 preparation histories of the head. Specific families of correlation 
 functions are more stable with respect to dissipation than the total 
 wave-function.
\end{abstract}

%%%%%%%%%%%%%%%%%%%%%%%%%%%%%%%%%%%%%%%%%%%%%%%%%%%%%%%%%%%%%%%%
% Introduction
%%%%%%%%%%%%%%%%%%%%%%%%%%%%%%%%%%%%%%%%%%%%%%%%%%%%%%%%%%%%%%%%
\section{Introduction}

It has been shown that 
certain computational problems scale 
more favorably when carried out on a quantum system than on {\em any} 
classical computer (see, e.g., \cite{EKE96}). The underlying ``quantum 
complexity'' may thus 
reduce computational complexity. Architectures for 
abstract quantum networks appropriate for such potential applications 
have been discussed (cf., e.g., \cite{BAR95}).

On the other hand, the control of such quantum networks appears to 
scale very badly with system size: In fact, in the language of 
statistical physics, that control would amount to use ``micro-states'' 
rather than ``macro-states'', a challenging undertaking, indeed. It 
should therefore not come as a surprise that virtually all proposals up to 
now face severe problems when trying to go beyond the (coherent) 
control of something like $N = 10$ pseudo-spins 
\cite{CIR95,DOM95,GER97,SHN97}. Also the detailed 
theoretical simulation will become increasingly difficult if not 
eventually impossible beyond that limit. Fortunately, 
as will be shown 
below, even such small networks may show a surprisingly rich behavior 
in terms of correlation-functions. Rather than the entanglement as such 
this pattern of correlations should be considered as the basis of the expected 
computational efficiency as well as other potential applications.

%%%%%%%%%%%%%%%%%%%%%%%%%%%%%%%%%%%%%%%%%%%%%%%%%%%%%%%%%%%%%%%%
% Composite systems
%%%%%%%%%%%%%%%%%%%%%%%%%%%%%%%%%%%%%%%%%%%%%%%%%%%%%%%%%%%%%%%%
\section{Composite systems}

\subsection{States} 
The system we are going to investigate here is 
composed of $M + 1$ spins, $\mu = S,1,2,\cdots, M$. The respective states are 
$|p(\mu )>,\, p=0,1$. 
The corresponding product basis is 
$|u(M)\cdots$ $r(2)q(1)p(S)> \equiv |u\cdots rqp>$. 
Arranged in the order of increasing binary numbers we also introduce the 
single-index notation $|s>$, $s = 0,1,\cdots, 2^{M+1}-1$ by identifying 
$|0>  =  |0\cdots 000>$, $|1>  =  |0\cdots 001>$, $|2>  =  |0\cdots 010>$, 
etc. This single-index representation will not only serve 
as a means to simplify some 
algebra. It reminds us that one can entirely avoid talking about entanglement 
while, nevertheless, keeping the product-space background still operative, 
though in a more subtle way: In terms of the {\em specific} operator 
combinations and their expectation values.

\subsection{Cluster-operators} 
For $M+1 = 5$ there are 
$(2^5)^2 = 1024$ orthogonal basis operators. One possible choice would be 
products of local transition-operators, 
$\hat P_{pq}(\mu ) = |p(\mu )~><~q(\mu )|$. For reasons that will become clear 
shortly it is more convenient to separate out 
the local unit operators $\hat 1 (\mu )$ so that the remaining operators 
become traceless. Such a scheme is provided by the Hermitian and unitary 
$SU(2)$-generators, $\hat\lambda_j(\mu )$,
\be
\ba
\label{lambda}
\hat \lambda_1(\mu ) & = & \hat P_{01}(\mu) + \hat P_{10}(\mu )\\
\hat \lambda_2(\mu ) & = & i \hat P_{01}(\mu) - i \hat P_{10}(\mu )\\
\hat \lambda_3(\mu ) & = & \hat P_{11}(\mu) - \hat P_{00}(\mu )\\
\hat \lambda_0(\mu ) & = & \hat P_{11}(\mu) + \hat P_{00}(\mu )\; 
= \; \hat 1(\mu)\,.
\ea
\ee
The corresponding product operators ($j,k,l,m,n$ = 0,1,2,3) \cite{MAH95}
\be
\label{q}
\hat Q_{jklmn} = \hat \lambda_j(S) \hat \lambda_k(1) \hat
\lambda_l(2)\hat \lambda_m(3)\hat \lambda_n(4)
\ee 
with $(\hat Q_{jklmn})^2 = \hat 1$ for any $(j,k,l,m,n)$ and
\be
\label{tr}
\mbox{Tr} \{ \hat Q_{jklmn} \hat Q_{j'k'l'm'n'}\} = 2^5 
\delta_{jj'}\delta_{kk'}\delta_{ll'}\delta_{mm'}\delta_{nn'}
\ee
then come in 6 classes, depending on the 
number $c=0,1,\cdots, M+1$ of subsystems they act on, i.e. the number of 
indices unequal zero. 
$\hat Q_{00000} = \hat 1$ is the only $c = 0$ cluster operator. When 
transcribed to the single index-space, $s = 0,1,\cdots, 2^5-1$, 
these operators appear like a set of ``generalized'' $SU(2)$-operators 
of the form given in eq.~(\ref{lambda}) with each single transition or 
projection operator replaced by a group of $2^M = 16$. 
Such operator combinations 
would be hard if not impossible to implement in a simple 
one-particle system with $2^5$ states; they reflect the structure of the 
underlying product space. Correspondingly, the expectation-value of 
any cluster-operator is a sum of $2^{M+1}$ density matrix elements in 
the single-index space. Examples for 
$M=2$ are the $c=1$-cluster operators,
\be
\ba
\label{q300}
\hat Q_{300} & = & (\hat P_{11} + \hat P_{33} + \hat P_{55} + \hat P_{77})
- (\hat P_{00} + \hat P_{22} + \hat P_{44} + \hat P_{66})\\

\hat Q_{030} & = &  (\hat P_{22} + \hat P_{33}+ \hat P_{66} + \hat P_{77})
- (\hat P_{00} + \hat P_{11} + \hat P_{44} + \hat P_{55})
\ea
\ee
or $c=2$-cluster operators such as,
\be
\ba
\label{q033}
\hat Q_{330} & = & (\hat P_{00} + \hat P_{33} + \hat P_{44} + \hat P_{77})
- (\hat P_{11} + \hat P_{22} + \hat P_{55} + \hat P_{66}) \\

\hat Q_{303} & = & (\hat P_{00} + \hat P_{22} + \hat P_{55} + \hat P_{77})
- (\hat P_{11} + \hat P_{33} + \hat P_{44} + \hat P_{66})\,. 
\ea
\ee

Any operator $\hat A$ in the $2^5$-dimensional Hilbert-space of 
spin-states can be 
represented as (summation over repeated indices),
\be
\label{a}
\hat A = \frac{1}{2^5}\, A_{jklmn} \hat Q_{jklmn}
\ee
with the parameters
\be
\label{ajkl}
A_{jklmn} = \mbox{Tr} \{ \hat A \hat Q_{jklmn} \}
\ee 
(Tr means trace over the total Hilbert-space.) 
In particular, the network-Hamiltonian $\hat H$ can be specified 
by the model parameters $H_{jklmn}$; they 
are usually constrained to 
$c$=0, 1 and 2-cluster-terms \cite{MAH95}. 
The density operator $\hat \rho$ is uniquely defined by the set of 
expectation-values (note that $\hat Q_{jklmn}$ is unitary)
\be
\label{k}
- 1 \leq K_{jklmn} = \mbox{Tr} \{ \hat \rho \hat Q_{jklmn} \} \leq 1
\ee
with $c$-cluster-operators defining $c$-particle correlations. 
For a pure-state, 
$\hat \rho = |\psi ><\psi |$, eq.~(\ref{k}) reduces to
\be
K_{jklmn} = <\psi |\hat Q_{jklmn}|\psi >\,.
\ee
By definition, $K_{00000} = 1$; 
the local Bloch-vectors $K_{j0000}, K_{0k000}, K_{00l00}$, etc. 
$(j,k,l=1,2,3)$ are equivalent to the respective reduced density matrices. 
A pure {\em local} state has Bloch-vector-length $1$. 
For so-called product-states all these correlations factor into one-point 
functions, i.e. $K_{jklmn} = K_{j0000} 
K_{0k000} K_{00l00} K_{000m0} K_{0000n}$, 
but, in general, they are independent. Local realism (cf. \cite{FER97}), to 
be sure, postulates that 
an appropriate distribution of local variables  (eigenvalues 
$\lambda_j = \pm 1$) could explain {\em all} these correlation functions 
rendering them statistically dependent; at least for larger networks 
this approach is no longer tenable. 
On the other hand, as will be shown below, the quantum mechanical 
evolution generates ``correlations between correlations''.

For later reference we also define symmetrized correlation-functions 
{\em within} one and 
the same system $\mu$: 
\be
C_{AB}^{(\mu)} = \frac{1}{2} (\mbox{Tr} \{ \hat \rho \hat A(\mu ) 
\hat B(\mu )\} + 
\mbox{Tr} \{ \hat \rho \hat B(\mu ) \hat A(\mu ) \} )\,.
\ee
Restricting ourselves to traceless operators, this correlation is independent 
of $\hat \rho$ (for two-dimensional Hilbert-spaces) and can simply be written 
as the normalized scalar product between the two representing vectors 
\cite{MAH95}; for $\mu=S$, e.g.,
\be
\label{cab}
C_{AB}^{(S)} = \frac{1}{2^{10}} \, A_{j0000}B_{j0000}\,.
\ee
 
\subsection{Unitary transformations} 
A unitary transformation of an operator $\hat A $,
\be
\hat A' = \hat U \hat A \hat U^{+}
\ee
with $\hat U^{+} \hat U = \hat U \hat U^{+} = \hat 1$, reads in terms of the 
$SU(2)$- parameters, 
\be
A'_{jklmn} = X_{j\; k\; l\;m \;n}^{j'k'l'm'n'} A_{j'k'l'm'n'}
\ee
where 
\be
\label{t}
X_{j\; k\; l\; m\; n}^{j'k'l'm'n'} = \frac{1}{2^5}\, \mbox{Tr} \{\hat 
U^+ \hat Q_{jklmn}\hat U \hat Q_{j'k'l'm'n'}\}
\ee
(For $\hat U = \hat 1$, $X$ is just the unit matrix, see 
eq.~(\ref{tr})). 
There are different types: 
We may distinguish transformations which operate in certain subspaces 
only. The locally selective transformation $\hat U(S)$ 
in the $n=2$-dimensional local Hilbert-space of $S$, e.g., is equivalent to 
a local rotation of the $SU(2)$-parameters with respect to the first 
index $j$, generated by (cf. \cite{SchLI95})
\be
\ba
\label{ts}
X_{j\; k\; l\;m \;n}^{j'k'l'm'n'} & = & X_{jj'}^{(S)} \delta_{kk'}
\delta_{ll'} \delta_{mm'} \delta_{nn'}\\
X_{jj'}^{(S)} &  = & \frac{1}{2}\, 
\mbox{Tr}_S \{ \hat U^{+}(S)\hat \lambda_j(S) 
\hat U(S) \hat\lambda_{j'}(S) \}\,.
\ea
\ee
(Here, $\mbox{Tr}_S$ means trace over the subspace of $S$ only.) 
As $X_{00}^{(S)} = 1,\; X_{jj'}^{(S)} = 0$ if either $j$ or $j'$ is zero, all 
parameters $A_{0klmn}$ are invariants. 
Correspondingly, a unitary transformation $\hat U(S,1)$ leaves the 
expectation values $A_{00lmn}$ unchanged, etc. These invariants (conservation 
laws) are important characteristics of the respective transformations.

\subsection{Time} 
As we do not consider equations of motion explicitly, 
time enters at most indirectly: To specify change, order and duration. 
For closed systems, unitary transformations are the only allowed type of 
changes (of states or observables) in closed 
quantum systems. Typically they are generated by the underlying Hamilton 
model. In the Schr\"odinger-picture this unitary transformation is applied to 
$\hat \rho$, 
in the Heisenberg-picture the inverse transformation (replacing $\hat U$ 
by $\hat U^{+}$ and vice versa) is applied to the observables.

Parameter time $T$ will come in with respect to the 
{\em order}, in which certain transformations are applied, as a 
{\em continuous 
parameter} controlling the individual transformation {\em quantitatively} 
(``pulse length'' $t$), and, eventually, with respect to the {\em order} of 
measurements. 
Finally, the induced dynamics can be characterized by correlation- 
and recurrence-times.

%%%%%%%%%%%%%%%%%%%%%%%%%%%%%%%%%%%%%%%%%%%%%%%%%%%%%%%%%%%%%
% The Turing model
%%%%%%%%%%%%%%%%%%%%%%%%%%%%%%%%%%%%%%%%%%%%%%%%%%%%%%%%%%%%%
\section{The Turing model}

Our system is sketched in Fig.~\ref{turing}: 
Spin $S$ is 
the ``Turing head'', the other $\mu =1,2\cdots, M=4$ subsystems denote 
memories (as part of a circular ``Turing tape''); the latter do not interact 
directly and are separated by ``empty '' cells. 
The head interacts with at most one cell at a time 
\cite{DEU85}; it moves clockwise and step by step to 
one of the $2M$ positions on the tape; 
there is no need for a feedback between the internal quantum state of 
the network and this pre-determined ``classical'' movement. 

We assume to have explicit control over the model parameters 
$H_{jklmn}$ defining the Hamiltonian, 
which  may even be modified in terms of pulses in parameter-time 
($ t_j$ is the pulse-length):
\be
\hat H (t) = \hat H_j \;\;\mbox{for}\;\;T_{j-1} \leq t < T_{j-1} + t_j
\equiv T_j\,.
\ee
Granted this access we can implement virtually any unitary transformation via 
\be
\hat U (T_{j-1}+t_j, T_{j-1}) \equiv\hat U_j = e^{-i \hat H t_j/\hbar}
\ee
($j$ is the step number), though this may seriously be limited in practice. 
If the Turing head is over an empty cell (tape position $2\mu -1$), a 
local transformation $\hat U_{\alpha}(S)$ on $S$ is applied, if 
it is in contact with a memory cell $\mu$ at position $2\mu$ a pair 
transformation on $(S,\mu)$ is induced ($\mu = 1,2\cdots, M$). 

\subsection{Local transformation on (S)} 
Let us consider the one-parameter-form 
\be
\ba
|0(S)> & \longrightarrow & \cos{(\alpha /2)} |0(S)> - 
i \sin{(\alpha /2)} |1(S)>\\
|1(S)> & \longrightarrow &  - i \sin{(\alpha /2)} |0(S)> + 
\cos{(\alpha /2)} |1(S)>
\ea
\ee
which can be generated by ($M = 4$)
\be
\label{us} 
\hat U_{\alpha } (S) = \hat Q_{00000} \cos{(\alpha /2)} - \hat Q_{10000}\;
i\sin{(\alpha /2)} = \hat U^+_{-\alpha }(S)\,.
\ee
According to eqs.~(\ref{us}), (\ref{ts}) and (\ref{q}), we find 
\be
\ba
X_{jj'}^{(S)}& = &\cos^2{(\alpha /2)} \;\delta_{jj'} + \frac{1}{2} 
\sin^2{(\alpha/2)} \mbox{Tr}_S \{\hat\lambda_1 \hat\lambda_j \hat\lambda_1
\hat\lambda_{j'}\}\\
&& + \frac{i}{4} \sin{\alpha }\; \mbox{Tr}_S \{\hat\lambda_1 \hat\lambda_j
\hat\lambda_{j'} - \hat\lambda_j \hat\lambda_1 \hat\lambda_{j'} \}
\ea
\ee
so that $X_{00}^{(S)} = X_{11}^{(S)} = 1,\; X_{22}^{(S)} 
= X_{33}^{(S)} = \cos{\alpha },\; 
X_{32}^{(S)} = - X_{23}^{(S)} = \sin{\alpha }$. (Here and in the following 
all terms not explicitly given are zero.) This matrix $X_{ij}^{(S)}$ defines a 
rotation of the Bloch-vector of $S$ around the $k=1$-axis in the 
$2,3$-plane. The phase $\alpha$ may 
be taken to result from a pulse of duration $t$
\be
\label{time}
\alpha = g t 
\ee
where $g$ would be the coupling strength to an external optical driving 
field. The correlation function between $\hat A = \hat \lambda_3(S)$ 
transformed by $ \phi$ 
and the same operator transformed by phase angle 
$\phi + \alpha$ then is, according to eq.~(\ref{cab}),
\be
\label{caa}
C_{33}^{(S)}(\phi , \phi + \alpha) = \cos{\alpha}\,.
\ee    
Based on eq.~(\ref{time}) this expectation value can be interpreted 
as a $2$-time $1$-particle correlation function in the Heisenberg-picture. 
Combinations of these have been shown to violate temporal 
Bell inequalities \cite{PAZ93}.

\subsection{Pair transformation on $(S, \mu)$} 
This unitary transformation is taken as the conditioned $\pi$-pulse, 
($q = 0,1$)
\be
\ba
\label{usmu}
\mbox{Resonance:}\; \;\; |0(S)0(\mu )> & 
\longleftrightarrow & |0(S) 1(\mu )>\\
\mbox{Off-resonance:}\;\; |1(S)q(\mu )> & \longleftrightarrow & |1(S) q(\mu )>
\ea
\ee
which we may write, in terms of cluster operators, in the form 
\be
\ba
\label{us1}
\hat U (S, 1)& = & \hat P_{00}(S) \hat \lambda_1(1) + \hat P_{11}(S) 
\hat 1 (1)\\
&=& \frac{1}{2}(\hat Q_{00000} + \hat Q_{30000} + \hat Q_{01000} - 
\hat Q_{31000}) = \hat U^+ (S,1)\,.
\ea
\ee
These operators $\hat U(S,\mu)$ commute; their implementation requires pair 
interactions, which make the transition 
frequency in subsystem $\mu$ depend on the state of subsystem $S$ 
\cite{MAH95}\cite{OBE88}. 
This transformation has become known as the (quantum-) 
controlled NOT \cite{BAR95}, as subsystem $S$ acts as a control for a 
$\pi$-pulse on $\mu$. We may associate a fixed pulse duration $t_0$ with 
this implementation; here we assume $t_0 \approx 0$. 
In general, the two types of unitary operators do not commute:
\be
[\hat U(S,\mu ), \hat U_{\alpha }(S)] = \sin{(\alpha /2)}( \hat 1(\mu ) -
\hat\lambda_1(\mu )) \hat \lambda_2 (S)\,.
\ee

%%%%%%%%%%%%%%%%%%%%%%%%%%%%%%%%%%%%%%%%%%%%%%%%%%%%%%%%%%%%%
% The process
%%%%%%%%%%%%%%%%%%%%%%%%%%%%%%%%%%%%%%%%%%%%%%%%%%%%%%%%%%%%%
\section{The process}

\subsection{The first cycle} 
We are now in a position to follow up the ordered sequence of $2M =8$ 
unitary transformations,
\be
|\psi^{(1,j)} > = \hat U_j |\psi^{(1,j-1)}>
\ee
where ($\mu = 1,2,\cdots, M$)
\be
\ba
\hat U_{2\mu - 1} & = & \hat U_{\alpha_{\mu}}(S)\\
\hat U_{2\mu} & = & \hat U (S,\mu)\,.
\ea
\ee
Here and in the following 
the upper index pair in parenthesis denotes the cycle number $m$ and 
the step number $j$, respectively. With $j = 2\mu$ ($\mu = 1,2,\cdots, M$) 
we may associate the time (cf. eq.~(\ref{time}))
\be
\label{T}
T_{2\mu} = \sum_{i=1}^{\mu} t_{2i-1} = \sum_{i=1}^{\mu}\alpha_i /g
\approx T_{2\mu - 1}\,.
\ee
$T_{2M}$ is then the time needed for each cycle. Now, 
let the initial state be $|\psi^{(1,0)} > = |0> = |00000>$ so that the local 
Bloch-vectors are given by
\be
K_{30000}^{(1,0)} = K_{03000}^{(1,0)} = \cdots = K_{00003}^{(1,0)} = - 1\,.
\ee
In the first step we apply the local 
transformation with a phase $\alpha_1$ leading to 
\be
|\psi^{(1,1)}> = \cos{(\alpha_{1} /2)}\; |0> - i\sin{(\alpha_{1} /2)}\; |1>\,.
\ee
In the second step we execute the pair transformation on $(S,1)$:
\be
|\psi^{(1,2)}> = \cos{(\alpha_{1} /2)}\; |2> - i\sin{(\alpha_{1} /2)}\; |1>\,.
\ee
In the third step we again apply the local transformation, now with phase 
$\alpha_2$, leading to 
\be
\ba
|\psi^{(1,3)}>& = &\cos{(\alpha_{1}/2)}\cos{(\alpha_{2}/2)}\; |2>
-i\cos{(\alpha_{1}/2)}\sin{(\alpha_{2}/2)}\; |3> \\
&&-i\sin{(\alpha_{1}/2)}\cos{(\alpha_{2}/2)}\; |1>
-\sin{(\alpha_{1}/2)}\sin{(\alpha_{2}/2)} \; |0>\,.
\ea
\ee
In the ``Heisenberg-picture'', this implies between step $2$ and step $3$ the 
local correlation as given by eq.~(\ref{caa}) with $\alpha = \alpha_2$. 
In the $4$th step the pair transformation on $(S,2)$ implies
\be
\ba
|\psi^{(1,4)}>& = & \cos{(\alpha_{1}/2)}\cos{(\alpha_{2}/2)}\; |6>
-i\cos{(\alpha_{1}/2)}\sin{(\alpha_{2}/2)}\; |3>\\
&&-i\sin{(\alpha_{1}/2)}\cos{(\alpha_{2}/2)}\; |1>
-\sin{(\alpha_{1}/2)}\sin{(\alpha_{2}/2)} \; |4>\,.
\ea
\ee
This procedure is continued with respect to the next memory cells $3$ and 
$4$ (steps $5$ through $8$). 
We note that the single-subsystem expectation values of subsystem $S$ 
and $\mu$ obey the relations
\be
\label{pro}
\ba
K_{30000}^{(1,2)}& = & - K_{03000}^{(1,2)} 
= K_{30000}^{(1,0)} \cos{\alpha_1}\\
K_{30000}^{(1,4)}& = & - K_{03000}^{(1,4)} 
= K_{30000}^{(1,2)} \cos{\alpha_2}\;\;\;\;\mbox{etc.} \\
K_{10000}^{(1,2\mu)}& = & K_{20000}^{(1,2\mu)} = 0
\ea
\ee
and as a consequence of the controlled-NOT-logic (cf. eq.~(\ref{q033})), 
\be
\label{anti}
K_{33000}^{(1,2)} = K_{30300}^{(1,4)} = K_{30030}^{(1,6)} = 
K_{30003}^{(1,8)} = - 1\,.
\ee
We thus see that the two systems, $S$ and $\mu$, are strictly anti-correlated
after step $2\mu$ 
(the state $|\psi^{(1,2)}>$, e.g., is actually an eigenstate of 
$\hat Q_{33000}$!), while 
the local Bloch-vector-lengths are less than $1$, i.e. local 
properties are not dispersion-free (``fuzzy''). This is typical for 
non-classical correlations. There can be {\em strict} correlations between 
{\em fuzzy} subsystems.

\subsection{Cycles $m \geq 1$.} 
We can summarize and generalize the above results by introducing the 
following functions: 
\be
\ba
\lefteqn{\kappa^{(m,2M)}(\alpha_1, \alpha_2\cdots, \alpha_j) = }\\
&&\frac{1}{2}[\cos{(m\alpha_1)}
\cos{(m \alpha_2)}\cdots \cos{(m \alpha_j)} ] 
+ \frac{1}{2}\left\{ \begin{array}{ll} 1 & m \;\;\mbox{even}\\
\cos{\alpha_1}\cos{\alpha_2}\cdots \cos{\alpha_j}& m \;\;\mbox{odd,}
\end{array}\right.
\ea
\ee
$\kappa_s^{(m,2M)}$ as above with $\cos{m\alpha_1}$ replaced by
$\sin{m\alpha_1}$, $\cos{\alpha_1}$ replaced by $-\sin{\alpha_1}$
and the $1$ replaced by $0$ ($j\leq M$), 
\be
\ba
\phi_k^{(m,2M)}& = &
 -\cos{(m\alpha_1/2)}\cos{(m\alpha_2/2)}\cdots
\cos{(m\alpha_M/2)} \hspace{1.8cm} m \;\;\mbox{even,}\\
\phi_k^{(m,2M)}&= & \cos{((m+1)\alpha_1/2)}\cos{((m+1)\alpha_2/2)}\cdots
\cos{((m+1)\alpha_k/2))}\\
&& \times \cos{((m-1)\alpha_{k+1}/2)}\cdots\cos{((m-1)\alpha_M/2)}
\hspace{1.6cm} m \;\;\mbox{odd,}
\ea
\ee
and $\chi_k^{(m,8)} = -\phi_k^{(m,8)}$ with $(m+1)$ replaced by 
$(m-1)$ and vice versa.
 
Then, at the end of each cycle $m$, the Turing head can be described by 
($M=4$)
\be
\label{ks}
\ba
K_{10000}^{(m,j)} & = & 0\\
K_{20000}^{(m,8)} & = & \kappa_s^{(m,8)}(\alpha_1, \alpha_2, \alpha_3, 
\alpha_4)\\
K_{30000}^{(m,8)} & = & - \kappa^{(m,8)}(\alpha_1, \alpha_2, \alpha_3, 
\alpha_4)
\ea
\ee
and the memory cells by 
\be
\label{km8}
\ba
K_{03000}^{(m,8)} & = & \phi_1^{(m,8)}\\
K_{00300}^{(m,8)} & = & \phi_2^{(m,8)}\;\;\;\;\mbox{etc.}\\
K_{03300}^{(m,8)} & = & \kappa^{(m,8)}(\alpha_2)\\
K_{00330}^{(m,8)} & = & \kappa^{(m,8)}(\alpha_3)\\
K_{00033}^{(m,8)} & = & \kappa^{(m,8)}(\alpha_4)\\
K_{03030}^{(m,8)} & = & \kappa^{(m,8)}(\alpha_2, \alpha_3)\\
K_{00303}^{(m,8)} & = & \kappa^{(m,8)}(\alpha_3, \alpha_4)\\
K_{03003}^{(m,8)} & = & \kappa^{(m,8)}(\alpha_2, \alpha_3, \alpha_4)\,.
\ea
\ee
The memory pair-correlations are all positive for $m$ even and 
decay with ``step distance'', i.e. the number of 
intermediate rotation and coupling steps to other memory cells 
(cf. also Fig.~\ref{tree}). 
The pair correlations between Turing head and the memories are given 
by
\be
\label{chi}
\ba
K_{33000}^{(m,8)} & = & \chi_1^{(m,8)}\\
K_{30300}^{(m,8)} & = & \chi_2^{(m,8)}\;\;\;\;\mbox{etc.}
\ea
\ee 

All the expectation values are strictly periodic in $m$ if 
$\alpha_j = 2\pi /p_j$ for all $j = 1,2,\cdots, M$ with $p_j$ a whole number. 
The period $p$ is then the smallest even number that has all these $p_j$ 
as factors.

In a similar way one obtains the results 
for step numbers smaller than $2M=8$. 
Generalizations to the situation where 
the phase angles differ from cycle to cycle are also straight-forward. 
For example, based on eqs.~(\ref{km8}),(\ref{chi}) 
we find a web of correlations like
\be
K_{33000}^{(m,j)} \cdot K_{03000}^{(m,j)} = K_{30300}^{(m,j)} \cdot 
K_{00300}^{(m,j)} = K_{30030}^{(m,j)} \cdot K_{00030}^{(m,j)}
\;\;\;\;\mbox{etc.}
\ee
valid for all steps $j$ within any cycle $m$.

%%%%%%%%%%%%%%%%%%%%%%%%%%%%%%%%%%%%%%%%%%%%%%%%%%%%%%%%%%%%%%%%
% Reduced descriptions
%%%%%%%%%%%%%%%%%%%%%%%%%%%%%%%%%%%%%%%%%%%%%%%%%%%%%%%%%%%%%%%%
\section{Reduced descriptions}
 
\subsection{Turing-head $S$} 
The description reduced to the subsystem $S$ is based on the local 
Bloch-vector 
$K_{j0000}, j =1,2,3$ only. Starting from the ground-state, 
$K_{30000}^{(1,0)} = -1$, this 
vector is subject to the rotation as given by eq.~(\ref{us}). We see that 
each controlled NOT operation implies a projection on the $3$-axis 
($K_{10000} = K_{20000} = 0$). 
The result of eq.~(\ref{ks}) for cycle $1$ 
is easily generalized to $M > 4$ with $\nu = 1,2\cdots, M$ and 
$\alpha_{\nu} = \pi / M$. We find
\be
K_{300\cdots}^{(1,2M)} = - \cos^M {(\pi / M)}\,.
\ee
With $ \alpha_{\nu} =  g t_{2\nu -1}$ (cf. eq.~(\ref{time}); $2\nu - 1$ is 
the step number), 
the quantum-Zeno-effect \cite{MIS77,KNI90,ITA90} results 
within the fixed time $gT_{2M}  = \pi$ (cf. eq.~(\ref{T})). 
It is interesting to note that the reduced density matrix (or Bloch-vector) of 
subsystem $S$ is, at any time $t$, identical with the density matrix of an 
{\em ensemble} of non-interacting spins (all with the same initial state 
and subject 
to the same local unitary transformation) but {\em actually measured} at each 
time 
$T_{2\nu}$, $\nu = 1,2\cdots, M$. For each ensemble member the series of 
measurements 
constitutes a ``decision-tree'', with each measurement result given by 
$K_{30000}^{'} = \pm 1$ (see Fig.~2). The ensemble average over 
these trajectories leads back to 
the behavior realized here by just one single object! The respective 
density matrices are identical. This is what one may call 
{\em quantum parallelism}. The interaction with the tape generates a 
dynamical evolution of the Turing head $S$ equivalent to $2^M$ different 
histories 
(cf. \cite{OMN92}), clearly an exponential gain. This will only hold, 
though, as long as no measurements are performed.

\subsection{Turing-tape} 
Contrary to the Turing head $S$, the other subsystems are each 
addressed by unitary 
transformations only once (within each cycle). Due to the built-in 
logic the state of subsystem $1$ 
is strictly anti-correlated with $S$ after preparation step $2$, 
subsystem $2$ is anti-correlated with $S$ after step $4$, and so on. 
This means that 
an actual projective measurement performed on these subsystems would reveal 
also the respective states of $S$. When the transformations 
are interpreted to happen in parameter-time $T_{2\mu}$, 
the subsystems $\mu \neq S$ indeed act as 
a kind of ``memory''. They allow delayed measurements on $S$. 
One may argue that 
this fact is the origin of the quantum-Zeno-effect discussed in Sect.~($5.1$): 
It suffices to {\em be able to measure} in order to get the 
freezing-tendency of measurements (``virtual watchdog''-effect). 

Local measurements of the memory cells amounts to the application of a 
projection- or transition-operator like $\hat P_{01}(\mu)$. As 
these operators commute among each other (for different $\mu$) and 
with any of the unitary operators {\em not} 
acting on $\mu$, we can postpone these measurements up to one cycle. For 
$\mu =1$, e.g.,
\be
\ba
\lefteqn{\hat U(S,4)\hat U_{\alpha_4}(S)\cdots \hat P_{01}(1)\hat U(S,1) 
\hat U_{\alpha_1}(S) |\psi^{(m,0)}> = }\hspace{2cm}\\
&&\hat P_{01}(1) \hat U(S,4)\hat U_{\alpha_4}(S)... \hat U(S,1) \hat 
U(S)_{\alpha_1} |\psi^{(m,0)}>\,. 
\ea
\ee
Let us first restrict ourselves to cycle $m=1$ with its decision tree 
(Fig.~\ref{tree}). The time order of 
these measurements (i.e. the measurement process) need not correspond to 
the time-order, in which the memory cells have been visited by the 
Turing head: The actual history for the latter (out of the 
possibilities as shown in Fig.~\ref{tree}) may thus be ``realized'' even 
backward in time! 

But not only this: The correlation between memory cell $1$ and $2$, e.g., 
must, by 
construction (cf. eq.~(\ref{anti})) and the invariance property 
$K_{03300}^{(1,8)} = K_{03300}^{(1,4)}$, 
reflect the correlation between the states of $S$ taken at $T_2$ 
and $T_4$, respectively. This is readily 
verified by comparing our result for $K_{03300}^{(1,8)}$, eq.~(\ref{km8}), 
with $C_{33}^{(S)}$ given by eq.~(\ref{caa}) 
(then a two-time correlation function in the 
Heisenberg-picture). The fact that $K_{03300}^{(1,8)}$ and $C_{33}^{(S)}$ 
are identical means, that a 
measurement of $K_{03300}^{(1,8)}$ can be used to infer the unperturbed 
$C_{33}^{(S)}(T_2, T_4)$. 
This holds, correspondingly, for 
$K_{00330}^{(1,8)}$ and $K_{00033}^{(1,8)}$. In this sense 
time-correlations of the past still ``coexist''.

As we continue into the cycles $m>1$, the unique identification 
of tape state and head history is gradually lost; histories 
become undecidable. The ``meaning'' of those measurements thus strongly 
depends on the step- and cycle number. 
At the end of cycle $m + p$, to be sure, the original 
situation is restored. The time-parameters $T_{2\mu}$ labelling those 
histories are thus defined only modulo $pT_8$ (if period $p$ exists).
 
%%%%%%%%%%%%%%%%%%%%%%%%%%%%%%%%%%%%%%%%%%%%%%%%%%%%%%%%%%%%%%%%
% Special machines
%%%%%%%%%%%%%%%%%%%%%%%%%%%%%%%%%%%%%%%%%%%%%%%%%%%%%%%%%%%%%%%%
\section{Special Machines}

\subsection{A ``coin-tossing machine''} 
For the machine defined by \{$\alpha_{\mu}=\pi /2; \mu = 1,2,3,4$\} 
all pair correlations and all 
one-point expectation-values are zero by the end of cycle $m=1$ 
(cf. eqs.~(\ref{km8}, \ref{ks})). The resulting histories all have the same 
probability and look like those of independent 
coin tossings at the times $T_{2\mu}$. As for the {\em ``Zeno-machine''} 
\{$\alpha_{\mu}=\pi /M; \mu = 1,2, \cdots ,M$\}, 
a complete measurement of the tape state at the end of 
cycle $m = 1$ would allow us to reconstruct the history of $S$. 
The period is $p = 4$.

\subsection{A ``cat machine''} 
As a next example let us consider the Turing machine defined 
by \{ $\alpha_1 = \pi/2, \alpha_2 = \alpha_3 = \alpha_4 =0$\}. 
The period is $p = 8$, again 
independent of $M$: $|\psi^{(m,j)}> = |\psi^{(m+8,j)}>$. 
At the end of any cycle $m$ all memory cells 
are strictly correlated (cf. eq.~(\ref{km8})). Furthermore, 
\be
\label{cat}
|\psi^{(1,8)}> = \frac{1}{\sqrt{2}}( |11110> - i|00001>)
\ee
is found to be a so-called cat-state, for which the decision tree of 
Fig.~\ref{tree} collapses to two histories only, $(1111)$ and $(0000)$, 
respectively. Moreover $|\psi^{(5,8)}>$ is a different one. As a process 
the built-up of these cat-states is thus quite simple. 
While cat states are reduced to product states by the decay 
(measurement) of any individual subsystem, all the memory pair 
correlations discussed here remain intact as long as the decaying 
subsystem is not part of that very pair. 
 
\subsection{Large-scale predictability} 
For $m = 100 < p$ and $M+1=10$ we would have roughly $m2^M \approx 
5\cdot 10^4$ 
transformations in a $2^{M+1}\approx  1000$-dimensional Hilbert-space; 
nevertheless, the calculation of these expectation 
values would scale, at most, linearly with $M$, independent of $m$! 
This indicates that simulations even of 
large networks could become feasible based on such rules. 
Of course, the number of expectation values 
increases exponentially with the system size $M+1$.

%%%%%%%%%%%%%%%%%%%%%%%%%%%%%%%%%%%%%%%%%%%%%%%%%%%%%%%%%%%%%%%%
% Conclusions
%%%%%%%%%%%%%%%%%%%%%%%%%%%%%%%%%%%%%%%%%%%%%%%%%%%%%%%%%%%%%%%%
\section{Conclusions}

We have discussed the dynamics of a special quantum network, 
which combines quantum-mechanical and classical features: 
The quantum-mechanical variables 
consist of a ``Turing head'' (pseudospin $S$) and a ``Turing tape'' 
($M$ memory spins). Classical variables are the phenomenological 
Hamilton-parameters, which are switched externally to generate 
discrete unitary transformations. 
The machine behavior is defined by its initial state and the phase 
angles $\alpha_{\mu }$ specifying those transformations.

This switching can be visualized as being induced by the 
Turing head performing pre-determined cycles over $2M$ Turing 
head positions. Correlations 
in terms of multi-point expectation values are built up in this 
process. Time defines the order of non-commuting operations and 
quantitatively controls transformation parameters.

The structure of these correlations may be attributed to 
the notorious ``holistic nature'' of quantum mechanics. Nevertheless, 
this built-up follows a strict logic; 
the type of admissible manipulations (rotations) is severely 
constrained in all but the simplest 2-level-space; this observation 
certainly applies to our present $2^{M+1}$-level-model. 
Additional constraints are built in by the 
selection of transformations which are actually implemented. Here they 
relate to the fact that the multi-levels actually refer to $M+1$ 
subsystems. These constraints are reflected by the spatio-temporal 
pattern of correlations.

There is probably good news and there is bad news as far as the 
consequences are concerned: 
The bad news is that the implementation of specific processes 
is much more constrained in 
the quantum regime than in the macroscopic world; this makes 
experimental progress in quantum computation depressively slow. 
The good news could be that, eventually, only constrained systems 
can make up a useful machinery; 
systems with large, unrestricted state spaces (like a free gas) 
are ``useless''. The 
constraints are something like fixed axles, wheels, and connecting 
rods in classical mechanics. Under fairly moderate conditions 
those correlations and the correlation between correlations should 
constitute a machine behavior. 
Rather than {\em enforcing} some specific 
behavior defined by abstract algorithms we might be better off trying to 
exploit the experimental repertoire of {\em real} quantum networks.

%%%%%%%%%%%%%%%%%%%%%%%%%%%%%%%%%%%%%%%%%%%%%%%%%%%%%%%%%%%%%%%%
% Acknowledgments and figure captions
%%%%%%%%%%%%%%%%%%%%%%%%%%%%%%%%%%%%%%%%%%%%%%%%%%%%%%%%%%%%%%%%
\section*{Acknowledgments and figure captions} 

We thank C. Granzow, A. Otte and R. Wawer for fruitful discussions.\\

\begin{figure}
\refstepcounter{figure}\label{turing}
\vspace{23cm}
\vspace*{-1.cm}%
\hspace*{-1.5cm}%
\includegraphics{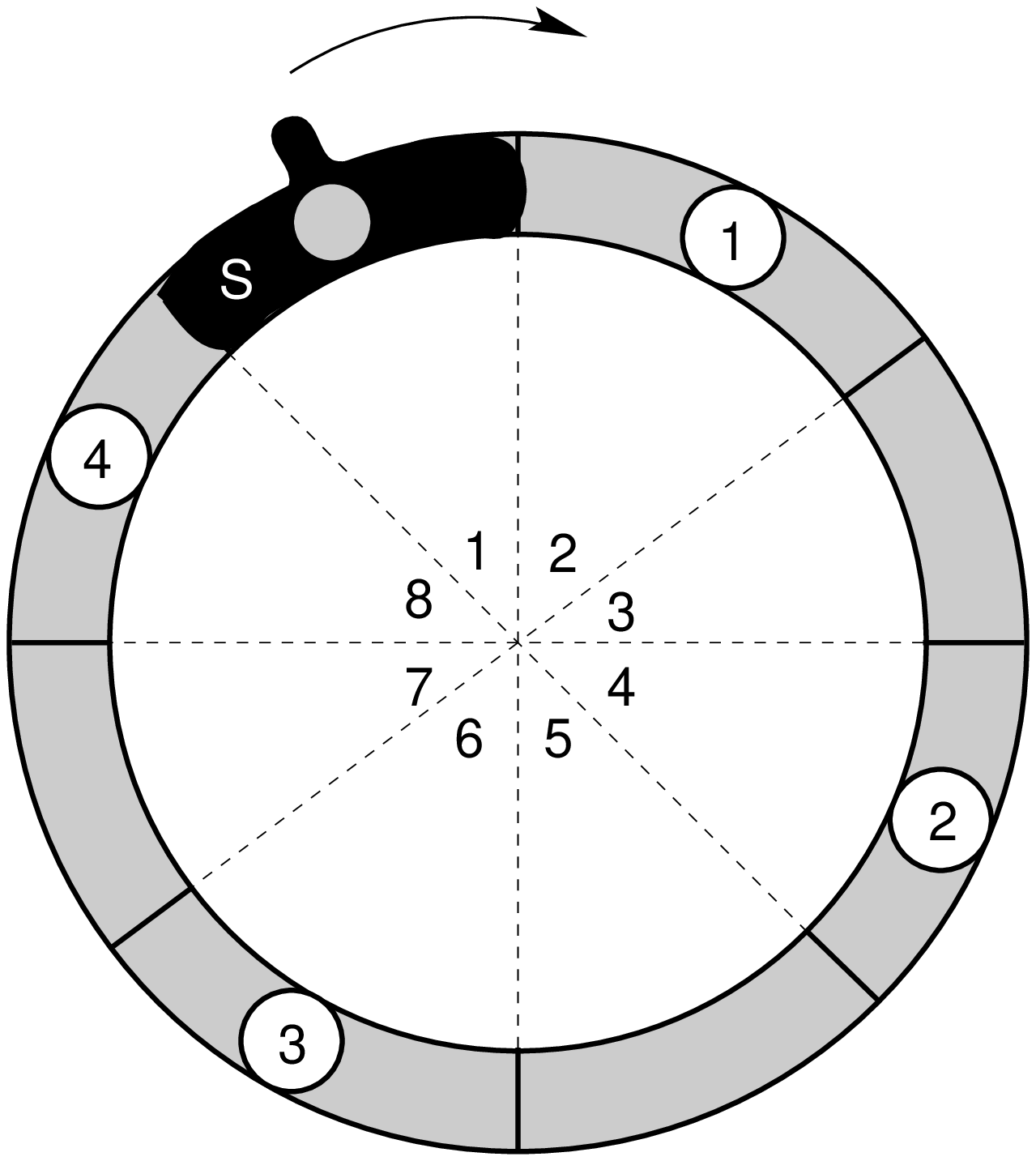}
\end{figure}

 Fig.~\ref{turing} Quantum Turing machine ($M=4$).\\ 
The circular Turing tape consists of $\mu = 1,2,\cdots, M$ memory cells 
(position $2\mu$) separated by empty cells (position $2\mu - 1$). 
The Turing head moves clockwise thus initiating a local 
(position-index odd) or a pair transformation, respectively (position 
index even).\\

\begin{figure}
\refstepcounter{figure}\label{tree}
\vspace{17cm}
\vspace*{-1.cm}%
\hspace*{1.cm}%
\includegraphics{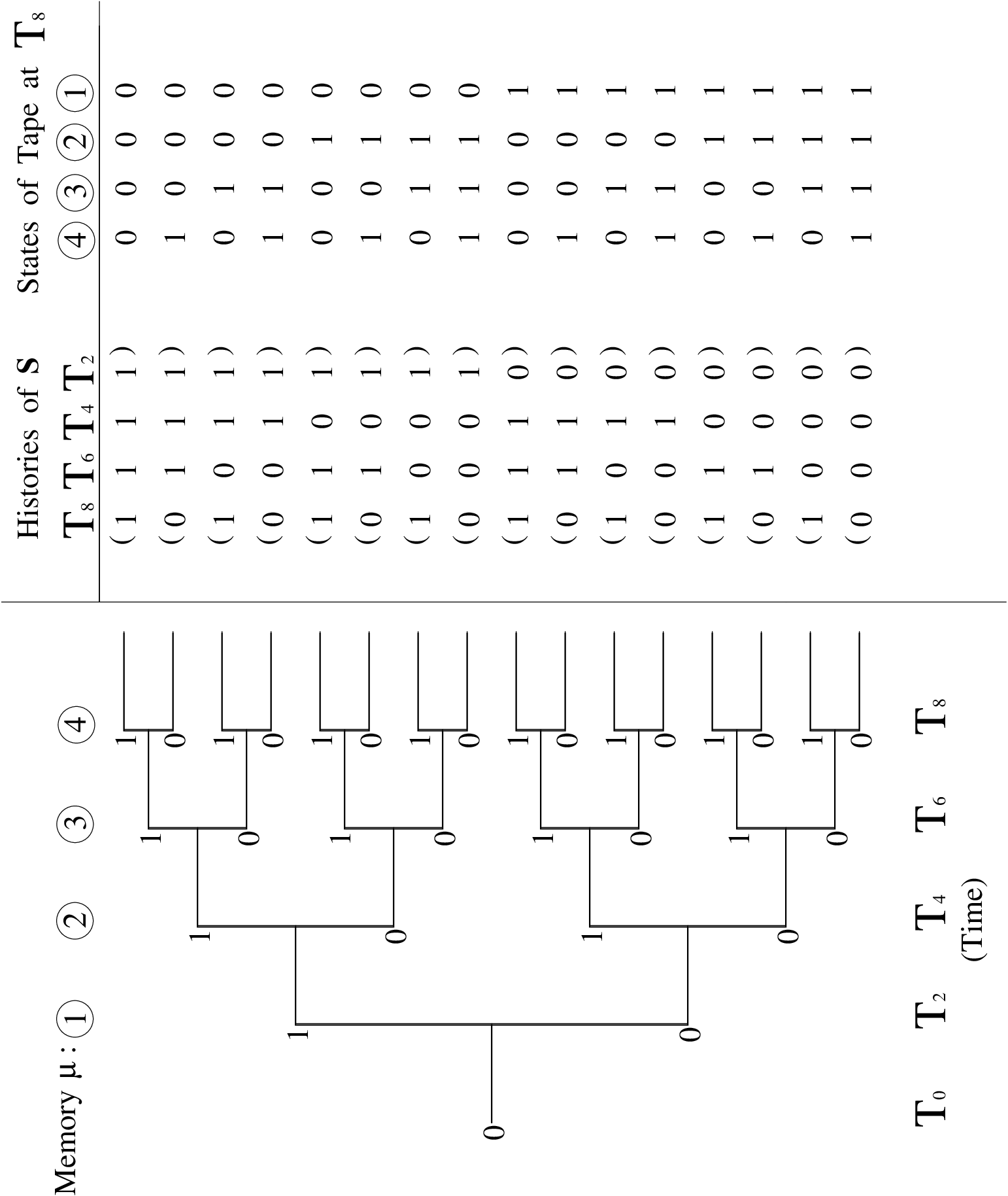}
\end{figure}

Fig.~\ref{tree} Alternative histories.\\ 
a. Decision tree with respect to step 
number $2\mu$ = 2, 4, 6, 8, as realized in an 
{\em ensemble} of non-interacting spins $S$ under the series of local 
transformations $\hat U_{\alpha_{\mu}}(S)$, but with immediate {\em actual} 
measurements (replacing $\hat U (S,\mu)$ of our Turing machine) 
at times $T_{2\mu}$. \\
b. For the single Turing machine all the possible histories are yet 
undecided and associated with the states of the Turing tape as given. 

%%%%%%%%%%%%%%%%%%%%%%%%%%%%%%%%%%%%%%%%%%%%%%%%%%%%%%%%%%%%%%%%
% References
%%%%%%%%%%%%%%%%%%%%%%%%%%%%%%%%%%%%%%%%%%%%%%%%%%%%%%%%%%%%%%%%

\end{document}